\documentclass[twocolumn,showpacs,preprintnumbers,amsmath,amssymb]{revtex4}

\usepackage{graphicx}
\usepackage{dcolumn}
\usepackage{bm}

\begin{document}


%

\title{Unification of Cosmology and Second Law of Thermodynamics: 
Solving Cosmological Constant Problem, and Inflation }

\author{Holger B. NIELSEN}
\affiliation{%
Niels Bohr Institute\\
17-21, Blegdamsvej,DK2100 Copenhagen\\
Denmark
}%

\author{Masao NINOMIYA}
 \altaffiliation[Also at ]
 {Okayama Institute for Quantum Physics, Kyoyama-cho 1-9, 
 Okayama City 700-0015, Japan.}
\affiliation{
Yukawa Institute for Theoretical Physics,\\
Kyoto University\\
Kyoto 606-8502, Japan
}%

\begin{abstract}
We seek here to unify the second law of thermodynamics with the 
other laws, or at least to put up a law behind the second law of 
thermodynamics.
Assuming no fine tuning, concretely by a random Hamiltonian,  
we argue just from equations of motion 
-- but {\em without} second law -- 
that entropy cannot go first up and then down again except with the 
rather strict restriction 
$S_{large} \le S_{small 1} + S_{small 2}$.
Here $S_{large}$ is the ``large" entropy in the middle era while 
$S_{small 1}$ and $S_{small 2}$ are the entropies  at 
certain times before and after the $S_{large} -$ era respectively. 
{}From this theorem of ``no strong maximum for the entropy" 
a cyclic time $S^1$ model world could have entropy at the most 
varying by a factor two and would not be phenomenologically realistic. 
With an open ended time axis $(-\infty, \infty) ={\bf  R}$ 
some law behind the second law of thermodynamics is needed if we  
do not obtain as the most likely happening that the entropy is maximal 
(i.e. the heat death having already occurred from the start).
We express such a law behind the second law 
-- or unification of second law with the other ones --
by assigning a probability weight $P$ for finding the world/the system in
various places in phase space.
In such a model  $P$ is almost unified with the rest as
$P = exp (-2 ~S_{Im})$  with $S_{Im}$ going in as the imaginary part of the 
action.
We derive quite naturally the second law for practical purposes, 
a Big Bang with two sided time directions
and a need for a bottom in the 
Hamiltonian density. 
Assuming the cosmological constant is a dynamical variable in the 
sense that it is counted as ``initial condition" we even solve in our model 
the cosmological constant problem \underline{without} any allusion to 
anthropic principle.
\end{abstract}

\pacs{11.10, 98.80.E, 98.80.C, 05.07}
\maketitle

\section{Introduction}
It is the purpose of the present article to make a sort of unification of
the second law of thermodynamics 
\cite{1}\cite{2}\cite{3} 
with the other laws in physics and 
cosmology, or rather we want to discuss how one may attempt to set up a law
behind this second law \cite{4}.
That is to say we want to make a formulation of a law at the macro physics 
level
that can lead to and explain the second law of thermodynamics.
Really we shall discuss such a potential law behind the second law of
thermodynamics being of the form of there existing a by fundamental law given
probability density $P(``path")$ assigning a probability density
-- in phase space -- to all possible solutions to the equations of motion,
called here abstractly ``path".
By assuming only mild generic properties of this probability density $P$
we then hope to get out not only approximately the second law of 
thermodynamics,
but also a bit of cosmological and physical results.
Most optimistically we believe our approach will play a role in solving the 
cosmological constant problem.

Some of the promising attempts to solve the cosmological constant
problem -- including the baby universe theory \cite{5}$\sim$\cite{8} 
attempt -- 
as to why the cosmological 
constant $\Lambda$ phenomenologically is so exceedingly small 
compared to an a priori fundamental scale for energy densities, 
using the Planck units as the fundamental units,
make use of getting this cosmological constant become rather a 
dynamical variable than a constant of nature only. 
For instance is the effective cosmological constant $\Lambda_{eff}$ 
corrected by influence coming in via the baby 
universes and thus has become dependent on ordinary dynamical variables.
In Guendelmann's solution \cite{9} to the cosmological constant problem the 
effective cosmological constant is also dynamical in the sense that 
it is some field expressions shown to be constant in both space and time
from what is essentially equations of motion or constraints.
Even in order to obtain the vanishing cosmological constant by means of 
anthropic 
principle \cite{Weinberg} it is also needed since the cosmological 
constant $\Lambda$ 
at least has the possibility of taking other values than just a single one.
Even in the derivation of the effective cosmological constant going to zero
as shown in the work(s) by R. Woodard and N. C. Tsamis \cite{10} 
there are such gravitational
fields around in what may look like the vacuum in practice that we in some
sense even in this model can claim that the effective cosmological constant
came to depend on fields and thus could be declared ``dynamical".

If indeed one has a model in which the status of the cosmological constant 
$\Lambda$ has become on an equal footing with dynamical variables 
such as fields or positions of particles, then the argumentation for
its value becomes in principle of the same nature as to argue about 
what the initial conditions for fields and particle positions are. 
Now it is the case since the separation between initial conditions and laws of
nature by Newton, that the usual mechanical laws and 
Maxwell equations etc. do \underline{not} tell about the initial
conditions. 
It is rather so that we define the ``initial conditions" to be the 
information which these laws for time development do \underline{not} 
deliver.
It is \underline{only} the second law of thermodynamics that says something 
about the initial state conditions, 
which otherwise are left over for purely observational determination.

{}From this logics it looks, that if we use as a solution of the cosmological 
constant problem one of 
the type in which $\Lambda$ is ``dynamical", we need 
to invoke some knowledge about initial state conditions.
Potentially we need some extension or model behind the second law of 
thermodynamics.

Now such a model behind the second law is at least a priori upset a bit by 
the famous problem of the arrow of time \cite{11}. 
Really the point is that mysteriously enough the second law violates 
drastically 
the time reversal symmetry!
which seems to be an at least approximate symmetry,
and indeed it holds to a very good approximation among the other laws in 
the Standard Model we could say.

This mysterious fact means that either we must live with one strange law 
not obeying time reversal invariance, and really not even 
CPT-symmetry, or we should have some sort of spontaneous breaking of 
time reversal invariance. 
Our cosmological models would indeed 
be strangely disturbed if we attempted to make second law be 
CPT-invariant i.e. for example, letting antiparticles have decreasing entropy.
Here we used the words ``some sort of spontaneous breakdown" 
to just mean that the fundamental laws have the symmetry in question 
-- time reversal or CPT -- but what goes on in Nature, 
the initial conditions, are not invariant under the symmetry.
Since usually initial conditions do not share the symmetries with 
the laws of nature there is no real reason that is against 
spontaneous breaking in the abstract sense mentioned above.
Also indeed it is one of the purposes of this article to 
present such a spontaneous-breakdown-like scenario for obtaining the 
second law.

At first glance it may seem that the Hartle-Hawking \cite{13} no boundary 
way of 
letting the universe development start looks like the model, we describe into 
our formalism from the fifth paragraph in section 2 below.This means 
essentially Boltzmann's
way \cite{14} which has a very special start.
However due to the imaginary time really used by Hartle-Hawking and the 
quantum mechanics inducing probability the Hartle-Hawking model for the actual
development of the universe may in fact come much closer \cite{15},
to picture of ours in the later part of the present article.
Our main unification analogy of letting the actual path of the development
be given by an imaginary part of the action may indeed be considered a 
generalization to a more general form of imaginary part of the action
than the one obtained in the special way of using complex or imaginary time
in the Hartle-Hawking scheme. \cite{12}

In the following section 2 we shall discuss unification of the second law
with the rest, the time development laws and introduce a probability
density $P$ over phase space as a replacement -- or behind-law --
for the second law.
In section 3 we put forward our attitude of avoiding fine tuning by 
thinking more concretely on a ``random Hamiltonian model."
Immediately we use this avoidance to produce a theorem restricting how
much entropy can grow and next diminish again. 
-- Of course such a discussion with diminishing entropy only makes sense
when second law is not assumed; but remember we seek to derive if rather --.
In section 4 we take the philosophy of a rather ``random" probability density
$P$ -- as introduced in section 2 -- and show that even without 
too much details we easily get suggestive arguments for the hope for 
``spontaneous-breakdown like effect."
A bit more detail is developed in section 5, where we stress the 
assumption of translational invariance. 
Finally we argue for an infinitely long time axis. 
Then in section 6 we study the asymptotic behavior in time.
Then in section 7 we discuss the benefit of having a short era with
lower entropy and in section 8 and we really get the second law out.

In section 9 we return to the asymptotics from section 7 and argue about 
the Hubble expansion having to go to zero for 
$t \to \pm \infty $ and thus solve the cosmological constant problem, 
provided the cosmological constant is assumed ``dynamical".

In section 10 we present some further results from our model: 
existence of approximate big bang and a bottom in the Hamiltonian density. 
In section 11 we conclude and come with some further outlooks.


\section{A Law behind the Second Law of thermodynamics}

Since as mentioned the second law seems somewhat outside the company 
of the other laws a unification of second law with the other laws seems 
a priori a special challenge.

It is clear that since the second law tells about initial conditions
or simply something about which solutions to the equations of motion are
realized -- at least in some statistical sense -- a law behind the second
law must be some law that assigns a probability to each of all the possible 
equation of motion solutions.
Since in a phase space description -- we here ignore quantum mechanics and 
work in the classical approximation in order not to invoke lots of 
further complications and philosophical troubles -- allows us to 
describe the solutions by their phase space point at say time 
$t=0$, a weight of probability for solutions could be described as a 
probability distribution over phase space destined to be used for the state
at time $t=0$, $(q(0), p(0))$.
Putting the existence of such a probability distribution over phase space at
a specific moment would be empty until we specify something about which 
weight to use. We can namely by such a weight describe any probability 
assignment to different solutions that would at all be possible.
Unless the second law were in some way logically contradictory it 
should therefore
also be possible to describe the knowledge somehow along such lines. 

It is the purpose of the present article to seek to unify the second law
with the other laws by imposing the symmetries of the other laws 
-- especially time translation invariance and time reversal invariance for 
example -- on the weight $P$ of probability function over the space of 
solutions to the equations of motion.
Such symmetry requirements are not an empty information input.
Rather it seems to severely endanger the chances of getting the second law out.

Even just imposing on $P$ that the probability of the different 
eq.\vspace{-.01cm} of motion 
solutions 
should depend on these solutions in a time translational invariant way seems 
a dangerous assumption. 
It would then seem that if we can conclude from a little bit of 
honey picked out of a bottle that there is high probability for being a lot 
more
honey in the bottle, then we might also think that honey tends to keep 
together 
in big bottles in the future and that by throwing some honey into a bottle
we can make it likely that it will contain a lot of honey.

A model that seems to roughly work in formulating the second law
-- but not caring so much for e.g. time translational invariance -- would be 
to take a model in which the phase space point in the ``first moment"
$t=t_{first}$, 
$(q(t_{first}), p(t_{first}))$
is used for the description.

Now we assume that we can in practice divide the phase space for the whole
world described by its micro degrees of freedom which are fields into various 
subsets called 
``macro states". 
Each macro state considered a subset should then contain those micro states
( = phase space points) that by some macroscopic observation could be 
distinguished as that macro state.
The macro state should then have to consider one state (from the macroscopic 
point of view.)

Now our expectations are in practice close to the following model:
In the early past i.e. $t=t_{first}$ there is no huge probability difference
between different macro states -- because they are all in practice 
``possible" -- thus if the volume $V(B)$ in phase space of a macro state $B$ 
is,
as is well-known, given as 
\begin{eqnarray}
V(B) = u_s^{2N}exp(S(B)),
\end{eqnarray}
we just need to, very crudely, have 
\begin{eqnarray}
P(\boldsymbol{q}(t_{first}), \boldsymbol{p}(t_{first})) \sim e^{-S(B)} ~,
 \nonumber\\
for~~~~~ ((\boldsymbol{q}(t_{first}, \boldsymbol{p}(t_{first})) \in B 
\label{star}~.
\end{eqnarray}

Here $u_s$ is a kind of discretization cutoff in phase space that
could represent Heisenberg's uncertainty relation, say
$u_s=\sqrt{h}$.
The number of degrees of freedom of the system is called $N$, and $S(B)$ 
denotes entropy
of the macro state $B$.

It is crucial for a good second law to come out that we do not assume the same
weight dependence at a later time, but only for $t=t_{first}$. 
This means that the proposed $P$-ansatz (\ref{star}) 
is very non-time reversal invariant.
At first it may seem hopeless to invent a time reversal invariant $P$ to
reproduce the violently non-time reversal invariant second law 
$\dot{S} \ge 0$ which under time reversal transformation changes totally 
sign and 
in the nontrivial case $\dot{S} > 0$ becomes directly contradictory with
itself.
Nevertheless we want in the present article to even argue that we can obtain
the second law {\em for practical purposes} from a $P$ that obeys a series of 
usual symmetries obeyed by the usual laws. 
Among these symmetries we even have time reversal invariance. 
The escape from the obvious contradiction is that we postulate below that 
in practice we only notice half the time axis so that under 
a strict time reversal inversion our world and the era would go into an era
``before Big Bang" which we do not in practice take seriously.
In other words we and our time would under time reversal go into a 
negative time $t < 0(= t_o)$ era which is not supposed to 
exist in usual thinking, but exists in our model; in this era only,
existing in our model 
one has the opposite second law, i.e. $\dot{S} \le 0$.
Even the time translation symmetry would be a threat of course if it were 
realized even 
as a symmetry of the state of the universe so that the state were constant 
since then $\dot{S} = 0$.
But here one just has to have that the solution realized does \underline{not}
have to share all the symmetries of the fundamental law, in this case 
e.g. $P$.  
That is a phenomenon much like spontaneous breakdown meaning that even
vacuum does not share the symmetries of the laws of nature,
so let alone the states of the universe.

To make the probability density $P$ as a function of the solution have a 
form compatible with the symmetry and locality properties of the other laws 
it is good to think of the analogy with the action $S(path)$ which is also
-- like $P$ a function of the path, the solution -- 
when it is extremized --.
The locality in time of the action means that it is additive in contributions 
from different regions in time. 
Really it is of the integral form
\begin{eqnarray}
S(path) = \int dtL
.\end{eqnarray}
This means in the Feynmann path integral 
$\int Dq e^{iS(path)}$ a factorization, so that we have for the integrand
$e^{iS}$ factors coming from the different time regions.
In order that effects of separate times should mean at first that one time 
could be treated independently of the other ones, the probability density
described by the  
path-dependent function $P(path)$ should be \underline{factored} into 
factors depending only on what  happens for the path in question in the 
different time intervals.
In this sense $P(path)$ is to be thought of as analogous to 
$e^{iS(path)}$.
Since we in quantum mechanics have the rule of squaring numerically the 
amplitude which in first approximation is $e^{iS(path)}$ having numerical 
square $|e^{iS(path)}|^2$, it is actually seen that $P(path)$ could be 
interpreted as an exponential of minus twice an imaginary part of the action
expressed by $S_{Im}$
\begin{eqnarray}
P(path) = e^{-2S_{Im}(path)}
.\end{eqnarray}

In general the locality in time meaning the just mentioned factorization of
$P(path)$ should mean that $P(path)$ should take the form
\begin{eqnarray}
P(path) = exp(-\int {\cal{P}}(\boldsymbol{q}(t), \boldsymbol{p}(t))dt)
\\\nonumber
= exp(-\int {\cal{P}}(\boldsymbol{q}(t), \dot{\boldsymbol{q}}(t))dt)
\end{eqnarray}
and we would even see the formal identification with imaginary part of 
the Lagrangian $L_{Im}$
\begin{eqnarray}
{\cal P} = 2L_{Im}
\end{eqnarray}
where $L$ stands for Lagrangian.

It must be immediately admitted that introducing such a probability weight 
$P(path)$ which depends on the behavior of the paths of all times is in
immediate danger of giving effects that make the model phenomenologically
non-viable. 
We postpone till section 12 the discussions and speculations as to 
how the immediate violation of observational facts is avoided by the 
natural requirements of Lorentz invariance and mass protection analogous 
to the Standard Model.
We can with the interpretation of ${\cal P} = 2L_{Im}$ claim a formal 
unification of
our ${\cal P}$, being a law behind the second law, and usual laws.

Imposing the same symmetries and locality restrictions as for the other laws 
means that we have to put for the expression for $\log P(path)$ the
same type of integral as we put for the ``real part'' of the action $S$.
For instance we would write down $\log P(path)$ in the general relativistic 
form imposing the usual locality and symmetry conditions as
\begin{eqnarray}
\log P = 
-\int \sqrt{g}\hat{\mathcal P}(\varphi, \psi, \cdots, \partial_\mu\varphi,
\cdots)d^4 x
\end{eqnarray}
where then the probability exponent density 
$\hat{\mathcal P}(\varphi, \partial_\mu\varphi, g^{\mu\upsilon},  
\partial_\rho g^{\mu\upsilon},\cdots )$
is a function of the fields and their derivatives at the point $x^\mu $
on the Riemann surface integrated over.

In the following sections we shall for the most part not use this detailed 
form 
of $P$
since we can mostly do with much more general considerations 
and we may thus
in principle have a better chance that the following considerations be true.

\section{Restrictions on entropy going up and then down}

It is the purpose of the present section to show without at all assuming to 
the second law some restriction on how the entropy 
can vary when the system passes through
the various
``macro states" just by 
using equations of motion and some avoidance of fine tuning. 
Really we use a philosophy of the Hamiltonian being under some restrictions 
to be considered random as our way of avoiding fine tuning. 
We shall indeed take it that the Hamiltonian or at least a part of it is a 
random function on phase space with the restriction, however, that it
(= the random part) does not cause the system to go from one macro state to 
another. 
That is to say the random part has zero Poisson-bracket with the macro 
variables, the 
variables characterizing a macro state.
Really our point of view is that we only accept generic solutions as 
truly existing.

Under the regime of this genericity philosophy we shall now show 
the following theorem (or lemma):

Theorem: Under the assumed random Hamiltonian setting: If a system 
passes successively in time and during a reasonable time through a series
of macro states among which three macro states have the entropies 
$S_1, S_2, S_3$ respectively
and are reached in this order $S_1, S_2, S_3$ in time, 
we have the restriction
\begin{eqnarray}
S_2 \le S_1 + S_3
\end{eqnarray}
alone from the equations of motion.

Here we introduced the somewhat cryptical notation of ``during a 
reasonable time". 
By this extra requirement is meant that the time intervals are supposed to 
be extremely small compared to numbers as $e^{S_1}, e^{S_2}$ and $e^{S_3}$
which describe the volumes of the phase spaces of the macro states in question.

Now the proof: Let us denote the biggest acceptable ``reasonable time" 
denoted by $T_r$
to be allowed for passage $T_r$. 
Then we have assumed 
\begin{eqnarray}
T_r \ll ~e^{S_1}, ~e^{S_2}, ~e^{S_3}
~.\end{eqnarray}
Really we even assume $\log T_r \ll S_1, ~S_2, ~S_3$.

We can now ask for the volume of the region inside the macro state with 
entropy $S_2$ which has any chance of being reached from any micro state
(meaning phase space point) in the first of the three macro states, namely 
the one with entropy $S_1$. 
If we denote by $v$ a typical ``velocity" of a point in phase space 
(It may be easiest to just take $v$ as unit and ignore if $v=1$), 
then the volume we asked for reachable from the $S_1$-macro state is 
estimated to 
\begin{eqnarray}
``Reachable ~volume" =  T_r\cdot v \cdot e^{S_1}u_s^{2N}
~.\end{eqnarray}

The $u_s^{2N}$ means a size for a Heisenberg uncertainty allowed phase 
space volume $u_s^{2N} = h^N$, where $h=2\pi \hbar$ is the Planck constant 
and $N$ is the number of degrees of freedom of the world-system considered. 
Really this $u_s^{2N}=\sqrt{h}^{2N}$ factor is only introduced to make 
dimensions be correct without giving the entropies very strange
dimensions. 
For simplicity of our arguments it is really best to put $u_s=1$.

Analogously we calculate the volume in phase space -- and it is also 
placed inside the macro state with entropy $S_2$ -- of all those micro 
states that can reach possibly the final macro state of the three ones, 
namely the one with entropy $S_3$,
\begin{eqnarray}
``reachable~~ part~~ of~~ macro~~ state~~ with~~ S_3" \nonumber\\
= T_r \cdot v\cdot u_s^{2N} \cdot e^{S_3} 
\end{eqnarray}
Now we shall remember that the volume of the part of the phase space 
corresponding to the macro state with $S_2$ as its entropy, the time-wise 
middle one, is 
\begin{eqnarray}
``Vol.~ of~ macro state~ with~ S_2" =  u_s^{2N} e^{S_2}
~.\end{eqnarray}

Then we use the philosophy
of random Hamiltonian to argue that this means
that the fraction of $u_s^{2N}$-volumes -- taken as a sort of cutoff 
in phase space -- that can be reached from the $S_1$-macro state is
\begin{eqnarray}
\frac{T_r  \cdot v \cdot e^{S_1}}{e^{S_2}} =
T_r \cdot v \cdot e^{S_{1}-S_{2}}
~.\end{eqnarray}

Statistically we expect therefore that there out of the
\begin{eqnarray}
T_r \cdot v \cdot e^{S_3} 
\end{eqnarray}
cells of size $u_s^{2N}$ that possibly can reach the $S_3$-macro state are
\begin{eqnarray}
(T_r \cdot v \cdot e^{S_3}) \cdot T_r \cdot \cdot v e^{S_{1}-S_{2}}
= (T_r \cdot v)^2 \cdot e^{S_{3}+S_{1}-S_{2}}
\label{success}
\end{eqnarray}
that come from the $S_1$-macro state. Since even the $\log (T_r \cdot v)$
is negligible compared to the entropies -- of course measure in 
Boltzmann constant as unit -- the condition for this number ($\ref{success}$)
to not to be much smaller than unity is that 
\begin{eqnarray}
{S_3}+{S_1}-{S_2} \ge 0
\end{eqnarray}
or
\begin{eqnarray}
{S_2}\le {S_1}+{S_3}
\end{eqnarray}
which were the condition we should prove. It is of course needed that 
at least one cell of the cutoff volume $u_s^{2N}$ should be able to come
through. 
End of the proof.

Formally this theorem seems to exclude the possibility of having both a 
big bang
and a big crunch with very low entropy and special states.
Such a scenario could at first seem quite attractive in the light of Hawking's
connection of the Hubble expansion time arrow and the second law of 
thermodynamics
arrow.
S. E. Rugh and one of us and later Hartle and Gellmann \cite{16} 
proposed a two-entropy picture for such a scenario
with both big crunch and big bang \cite{17}\cite{18}, 
but from the present theorem 
this kind of scenarios seem to have trouble.

\section{Time-axis discussion}
If we make a very strong assumption of time translation invariance we are
forced to consider the set of all moments of time -- the time axis 
so to speak -- to be either the set of all real numbers ${\bf R}$ or 
a compactified circular ``time axis" isomorphic to the $S^1$ circle. 
If we have the perhaps phenomenologically more realistic half interval 
with a certain starting time (creation), Big Bang, we do not strictly 
speaking have time translational invariance.
With this excuse we shall only treat the mentioned two possibilities, 
${\bf R}$ and $S^1$. 
We shall quickly exclude $S^1$ and the model put forward in the present
article as our viable model is one with time axis ${\bf R}$. 
We thus strictly speaking have the problem of needing a bouncing back 
Big Bang instead of having the singularities needed by the Hawking-Penrose 
singularity theorem. 
In the light of the fact that we do not really know quantum gravity this 
possible
trouble by needing a bounce should not be considered a so extremely
severe problem, especially not in the very abstract formulation used in 
the present article. 
Let us first get over the possibility of the compactified time forming an 
$S^1$-circle, i.e. a world with an intrinsic period $T$ say imposed. 
In such a world the periodicity requirements are in number equal to the 
number of ``initial state" describing variables, namely twice the number 
of degrees of freedom, the dimension of phase space. 
This means that in the generic philosophy --random Hamiltonian -- there 
will only be a discrete set of solutions to the equations of motion.
This means that there is, apart from the discrete possibilities, a unique 
solution, so that any extra assumptions about ``initial conditions" become
in first approximation superfluous. 
So long as we consider the solution essentially uniquely determined there
would in the $S^1$-circle case be no place for the in section 2 introduced 
probability weight. 
In the $S^1$-circle case we could therefore at best hope for the second law 
of thermodynamics simple to come out by itself in this approximation, without 
any behind-law
being put in. 
It also looks in the first go as if the compactified time axis is indeed
helpful to get in the direction of the second law, in as far as 
we indeed obtained 
that the entropy in compact time world model is about equally likely to 
different entropy values.
That is to say that the periodicity requirement is very likely to enforce
the likely solution to have less entropy than maximal
entropy. 
This is already to be considered a step in the right direction in as far as
a random point with the natural phase space measure as probability 
distribution (i.e. $P=1$ in our notation of section 2) 
would with very high probability be in the one of the macro state(s) with 
the maximal entropy -- the heat death macro state. -- 
Therefore already the fact that one does
not get the heat death as prediction means that the compact world model
even without any P-probability density imposed provides the answer to what
is already a mystery:  Why is universe not, now say, in the heat death 
state?
However we found that the $S^1$-compact time world also leads to the entropy
being totally constant as a function of time all the circle around. 
That there is at least strong limits to how much entropy could possibly
vary in compact time or intrinsic periodicity model can be seen by applying
our theorem in section 3.
Since the time is on the circle we can in fact always choose to consider
the highest entropy $S_{h}$ reached in time in 
between the lowest $S_{l}$ on both sides.
Thus our restriction from the theorem will tell us
\begin{eqnarray}
S_h \le 2S_l
\end{eqnarray}
so that alone from this theorem the highest entropy $S_{h}$ reached
along the time-circle could not possibly be more than at most twice the 
lowest one $S_{l}$ reached.

However, if we reconsider the proof of our theorem above with the 
compactified time $S^1$ in mind we easily find that the chance for 
finding at least one cell of $u_s^{2N}$ size in the macro state with the
highest entropy $S_{h}$ that comes from \underline{and} goes to 
the same micro state cell in the low entropy $S_{l}$ macro state is 
crudely by ignoring factors $T_r^v$ etc.
\begin{eqnarray}
e^{S_l - S_h}
.\end{eqnarray}
The equation of motion development backward and forward from the 
$S_{l}$-entropy macro state have to ``accidentally" meet inside the 
$S_{h}$-entropy macro state phase space volume $u_s^{2N}e^{S_h}$.
Thus the generalization of our theorem to the compact time case really leads to
\begin{eqnarray}
S_h \leq S_l
\end{eqnarray}
which of course implies that entropy must be totally constant $\dot{S} =0$ as
we already claimed in \cite{19}.

Phenomenologically we have often $\dot{S} >0$ strictly and so that $S^1$-time 
system does not fit to this \underline{non-trivial} realization of the
second law of thermodynamics.

Keeping to the assumption of so strong time translational invariance that
we even shall be able to find a moment on the time axis time translated by an 
arbitrary  time-distance relative to any time-moment we are therefore 
driven to the time axis being the set of all real numbers.

In such a world -- with the time axis $(-\infty, +\infty )=
{\boldsymbol{R}}$ --
we shall certainly not get anything like second law without imposing a 
drastically varying phase space probability density $P$.
Indeed if we look $P \sim 1$ we would get the heat death prediction
which is a priori the first expectation.
With the two infinite time direction states left free we do not
as in the $S^1$-time axis case have any restrictions that can prevent the 
maximal entropy $S_m$ or heat death macro state.

In the ${\boldsymbol{R}}$-time axis model --which we shall use in the bulk 
of this
article -- we thus need the probability density $P$ law as law behind the 
second law of thermodynamics.

\section{The behavior for time ~$t \to \pm \infty $}\label{secV}
Let us now consider the infinite time axis model with some probability 
density $P$.
Even if some entropy variation occurs one will expect asymptotic behaviors
for $t \to \pm \infty $, which we should study first,i.e. in this section.

We have already put forward a concept of macro states to describe 
subregions of the phase space in which at least to some degree the 
system can remain for some time. 
However some of these macro states may be much more stable than others. 
Potentially we might find classes of macro states which could be counted
together and function together as a more stable macro state in which the 
systems stays longer.
In the asymptotic times we would expect to find the system/universe in a 
class of macro states which is very stable.
We must with a probability weight really expect that in the $t \to \pm \infty $
time regions the stable class of macro states chosen by Nature must be the
one with on
the average the highest value of $<P\cdot e^S>$.
By this $e^S$ in our a bit formal expression $<P e^S>$
we mean the phase space volume, really it is $u_s^{2N}e^{S}$, for
the set of states counted to the macro states together in the class of macro 
states
that together were stable.
The average sign $<\ldots >$ means that we take the average over associated 
micro states in the macro state(s).

We expect that the macro variables in such a most stable class of macro states
are either moving periodically or have basically stopped moving.

As one of the major macro variables we can phenomenologically think of the
size $a(t)$ of the universe as it occurs in usual Robertson-Walker
cosmological model. 
If as we have just suggested the $<P e^S>$ being maximized determines the 
asymptotic values or oscillation intervals for the macro variables,
we might expect also that
the universe size variable $a(t)$ would in the asymptotic stop.
That should mean that Hubble expansion should essentially stop asymptotically.

Phenomenologically it is at least true that compared to the suspected 
fundamental scale for time, the Plank time $10^{-43}s(second)$,
the Hubble constant $H$ is already exceedingly small compared to the inverse
fundamental time $10^{43}s^{-1}$.
In this $-a$ bit thin -- way we can claim that the universe already develops
very slowly.

But why should the universe not continue to inflate strongly
(say on Planck scale), since that would just make the density even lower and 
the stability even better in later times?
This could be explained if there -- with a 50\% accident --
were a term making $\log P$ fall as the universe radius grows.
Then namely it would not ``pay" in terms of $<P>e^S$ to let the universe grow
faster than needed to satisfy enough stability.

\section{Deriving second law of Thermodynamics}
As we shall argue for in section \ref{secVII} 
below, it is very likely that there 
shall be a short era with very small entropy. Then there be transition times 
between this era and the asymptotic eras described in section \ref{secV}.
 
Our major point is that in these transition times we will 
-- almost certainly -- have variation of the entropy.
Indeed to find the for the short time interval to be used state(s)
will most likely mean finding and realizing a very small phase space
region meaning a low entropy macro state.
One can easily imagine that by some accident certain low entropy state is
a very high $P$ state.
We will therefore very likely find that in the time direction away from 
the time, $t_0$ say, when the huge $P$ but ``unstable" macro state(s) is/are
realized the entropy will increase.
That is to say:  
\begin{eqnarray}
\dot{S} > 0, ~~for~~ t > t_{0} 
\nonumber\\
\dot{S} < 0, ~~for~~ t < t_{0}
\end{eqnarray}
If indeed -- as is very likely -- the entropy around the time $t_0$ is very 
small
compared to entropies achieved at other times, our theorem in section 3 will
essentially come to mean that on each side of $t_0$ the time axis we must have
that entropy becomes a monotonous function of time.

Formally this is very easily seen by using the very likely happening that 
$S(t_0)$ very small.
Then two times later than $t_0$, say
\begin{eqnarray}
t_0 < t_1 < t_2
\end{eqnarray}
we use our theorem to derive
\begin{eqnarray}
S(t_1) \le  S(t_2) + S(t_0)
\end{eqnarray}
which for $S(t_0)\ll S(t_2), S(t_1)$ means
\begin{eqnarray}
S(t_1) \lesssim S(t_2)
.\end{eqnarray}
That is to say that in the region $t > t_0$ 
we have increasing entropy
\begin{eqnarray}
\dot{S}(t) \ge 0  ~~~for~~~ t\ge t_0
.\end{eqnarray}

Similarly we see again by using our theorem and the approximation that 
$S(t_0)$ is also very small compared to the entropies in the earlier than
$t_0$ times $t < t_0$, i.e. that
\begin{eqnarray}
S(t_0) \ll  S(t)
\end{eqnarray}
also for $t<t_0$, that if
\begin{eqnarray}
t_{-2} < t_{-1} < t_0
\end{eqnarray}
then
\begin{eqnarray}
S(t_{-1}) \le  S(t_{-2}) + S(t_0) \approx S(t_{-2})
.\end{eqnarray}
Thus in the $t < t_0$ times we have
\begin{eqnarray}
\dot{S}(t) \le 0 ~~for~~ t<(t_0) 
\end{eqnarray}
i.e. falling entropy.
Of course we would just invert the time axis and claim as usual that we
have increasing entropy and 
thus that usual second law holds also if we happened to live in the 
era $t<t_0$.

In this sense we have derived as almost unavoidable that in a model with a 
strong probability density variation over phase space the most likely,
in the $P$-sense, solution(s) to equations of motion will effectively look 
as having the second law, provided though the following small caveats of 
interpretation;

1) We shall choose the time axis so that positive $t$-direction is
counted away from the $t_0$-time. 
We redefine for this purpose a new time $t_{new}$ as 
\begin{eqnarray}
 t_{new} &=& t-t_0 ~~for~~ t>t_0
\nonumber\\
t_{new} &=& t_0 -t ~~for~~ t>t_0
~.\end{eqnarray}

2) We only derived the $\dot{S}=\frac{dS}{dt_{new}} \ge 0$ for one half
of the time axis. 
To have it -- the second law -- we must ignore what goes on on the other side
of the time $t_0$.

3) Our derivation only worked under the though suggestive assumption that we
could argue for -- or generally expect -- that the entropy enforced at the
time $t_0$ was indeed so small that it were negligible compared to the other 
entropies.
Really though this seems to be almost unavoidable a correct assumption, 
but it might deserve further theoretical study.

Phenomenologically of course the time $t_0$ is being identified with the
Big Bang time. 
But notice then that our model suggests that there be a bounce from a 
contracting universe, though one with opposite second law.

\section{Do we obtain Big Bang?}\label{secVII}
In the light of that it is unavoidable with essentially, whatever
the probability density function  $P$  might be, that we should have the 
universe  behaving corresponding to highest   \quad
$<Pe^S>$ stable class of 
macro states, we might now -- somewhat worried from phenomenological point
of view -- ask whether the totally most favoured, most probable with $P$,
development would not be to have the approximately static universe all 
through from -- $\infty $ to $+\infty $ in time.

The answer, we shall show is that such an all the time like the asymptotic
region behaving universe is \underline{not} likely to be the most likely 
outcome. 
The point is that we can very easily risk that it would pay 
probabilitywise
to get even for limited time interval the system in a state with a huge 
${\cal P}$ 
compared to the asymptotic average.
Since we in seeking the macro states for the asymptotic only could look for
the subclass of ``stable" classes of macro states, it is indeed very likely
-- almost surely -- so that there exist ``unstable" macro states outside the 
usable classes of stable collections of macro states for which the 
$<Pe^S>$ is much higher than for any ``stable" competitor.
If this is so a more likely universe development than the one being in the
``Stable class" all along the time axis would be one in which the more
high $Pe^S$ macro state is reached during some finite time while we still
have the already described asymptotic behaviors.
Then of course there is needed on both sides on the time axis some development
between the ``unstable" utterly special $P$-favoured state and the asymptotic
behaviors.

\section{The cosmological constant problem}
As mentioned in the introduction a good start on solving the cosmological
constant problem is to make the cosmological constant become a dynamical
variable on an equal footing with the fields or positions of particles.
However, logically this completely alone cannot be the whole story for 
even if the cosmological constant is in some model part of the dynamical 
variable
of the 
system, why should it be extremely small, almost zero, for that reason?
It is at this stage that our present work can have a role to play.
We shall thus simply assume that by some mechanism on the market 
we have got to the effective cosmological constant being a function of
variables in the phase space of states of the universe.
Or we could let it, the cosmological constant, simply be one of the 
variables, one of the generalized coordinates.
We must of course then in some way or the other assume or obtain that 
it be a constant in time.
That is to say we must somehow achieve $\dot{\Lambda}=0$
where $\Lambda $ is the cosmological constant.
However, it may well be allowed that our probability density function $P$
can depend somewhat on $\Lambda $, and most importantly the value of
$\Lambda $ has great significance for the development of the other dynamical 
variables in the model.

We already argued in section 6 that the Hubble expansion should end up 
being small so as to lead to an asymptotic state or class of macro states.
Since ignoring matter one has in gravitationally determined units
\begin{eqnarray}
H^2 \approx \Lambda 
\end{eqnarray}
a small $H$ already suggests a small $\Lambda $.
This was, however, only true ignoring matter.
There are however reasons to believe that requiring the ``stable" macro states
or classes of them called for in section 6 the universe does the best job of
stability by having an exceedingly low mass density.
In the very long run the formation of black holes could be an important 
mechanism for producing entropy since by means of them you can liberate even 
the free energy contained in the baryon masses.
But if it happens that the large black hole states are not so favoured 
with respect to $<Pe^S>$ they should be avoided in the asymptotic most 
favoured stable states.
That might be achieved by having low matter density, because it then
simply becomes impossible for enough matter to collect at one spot 
so as to form black holes.
In this way we may understand or derive that stability against falling into a 
possibly less high $<Pe^S>$ state with black hole(s) will require a low
matter density.
Thus we may see that the asymptotic state which should be the 
most favoured $<Pe^S>$ among the stable states must be a low matter density one
since otherwise it would not be ``stable".

But then we see that both matter density $\rho_m$ and Hubble expansion
rate must be small in the asymptotic state.
Thus we see that indeed with a ``dynamical" cosmological constant it must
be small.

This we can consider a solution to the cosmological constant problem,
but we must admit of course that we only succeeded under the assumption 
that the cosmological constant were influencable at all.

On the other hand we must also remark that it looks extremely hard to give 
an explanation to the smallness -- of what we really have to -- of the 
effective, or dressed, cosmological constant without any chance that 
it is influencable. 
Precisely because it is the dressed or effective cosmological constant 
rather than the bare one that should be zero or small, it depends
on all the vacuum energy density corrections.
It is very hard to see how all these -- huge-- corrections should be just
cancelled by terms in the theory for the bare cosmological constant unless it 
can
become influenced somehow from the vacuum having all the contributions of
different forces.
Thus it is on general grounds extremely hard to see, how one could get the 
cosmological constant problem solved 
if the cosmological can not be influenced some way or an other,
unless somehow the corrections to the vacuum energy density
is zero or some very simple value. 
This latter is the possibility of global supersymmetry in which 
the corrections are zero and also the cosmological constant.
But as is well known the pure global SUSY explanation for the vacuum energy
density cannot solve the cosmological constant problem with sufficient 
accuracy because the SUSY-breaking needed for phenomenology is much bigger
than the actual value for this constant.

\section{Inflation and bouncing scenario}
It is actually very natural and easy to obtain in our model an inflation
scenario \cite{inflation} 
-- as is well known to be strongly called for phenomenologically --
because the function $\mathcal P$ in terms of which a suggestive form for 
the probability
measure density $P$ is written 
can easily be a function of some scalar field $\varphi (x)$, 
the inflaton field. 
It would then be extremely likely because $P$ would be big to have a period
era in which the various scalar fields would take values just favouring
$\mathcal P$ to be as small as possible and the $P$ large.

Such an inflation with $\varphi$ near the low $\mathcal P$ value would normally
not be stable but the inflation field would roll down and ``re"heating
\footnote{Since it is a peculiarity of our model that from $t_0$ 
on it was cold all the time during inflation ``reheating" as one usually
calls the heating after inflation is strictly speaking a wrong terminology.}
would occur causing a creation of a lot of entropy.
With the bit of adjustment of cosmological constant and the expansion rate
resulting we could get approximately the asymptotic conditions called for to
have a ``stable" class of macro states.

With the $\mathcal P(\varphi(x)_{\ldots} )$ being taken the same all over 
space time
would make the same inflation starting value $\varphi(x)=\varphi_{start}$. 
So the inflation scenario is rather a simple inflation picture homogeneous
from the start than a chaotic inflation picture.

An interesting point that turns out also to concern inflation is the 
prediction above that we get two half time axis worlds only separately having 
a second law. 
This picture calls for a rebounce scenario rather than a big bang
singularity.

On the other hand if we let the $Pe^S$ favoured state that so to speak has
to be realized -- around the moment we call $t_0$ -- be the 
inflating state, then there would be no call from $Pe^S$ for a pre-inflation
state of the universe.
Just inflation would be enough.

Luckily enough for the working of our model:  In spite of that it is hard to
get rebouncing -- Penrose-Hawking theorem -- a model dominated by an
inflation state just with an essentially constant scalar field
-- or several scalar fields -- would simply be a de Sitter model.
A de Sitter universe actually does precisely rebounce provided the universe
has a positive spatial curvature so that $K=+1$.
We have namely the Einstein equation in the Friedmann model:
\begin{eqnarray}
(\frac{\dot{a}}{a})^2 = H_0^2 
\left(\frac{\rho}{\rho_{crit}^{0}}-\frac{k_0}{a^2}+\lambda _0\right)
\end{eqnarray}
showing that as $a$ approaches $\sqrt{\frac{k_0}{\lambda_0}}$ the
Hubble expansion $H=\frac{\dot{a}}{a}$ goes to zero.

Here $\lambda_0$ is given by
\begin{eqnarray}
\lambda_0 &=& \frac{\Lambda c^3}{3H_0} \nonumber\\
H_0 &=& \left(\frac{\dot{a}}{a}\right)_0
\end{eqnarray}
where the suffix $0$ means the value at present time.
\begin{eqnarray}
\rho_{crit}^{0} = \frac{3c^2}{8\pi G}H_0^2
\end{eqnarray}
is the critical energy density. 
The curvature parameter $k_0$ is given by $k_0 = \frac{Kc^2}{H^2}$.

When both $k$ and the cosmological constant, which in gravity specified 
units is $\lambda_0$, are positive, the bounce, $H=0$, occurs
for a finite real value of the radius $a$ of the universe.
So if we take it that infinite universes with $k=-1$ or $k=0$ are
``unpleasant" philosophically we will get very nicely a rebounce in the 
inflational time itself.
This liberates our model from a potentially great problem: rebounceing
could be very difficult to achieve.

\section{Further results}
We should here stress that our model in addition to the major results already
mentioned: the cosmological constant being small and the effective second
law, for the half time axis in which we live, further leads to:

A) The existence of an approximate big bang in the sense that the universe
once were very much smaller than today.
We find, however, not a true starting singularity but rather a 
rebouncing in the inflational period. 
There is no singularity around the 
time $t_0$ (of course in usual cosmologies taken as $t_0=0$), but rather that
an $S^3$-spherical universe with at that moment an effective positive
cosmological constant due to some inflation field(s) rebounce smoothly.
The Hubble constant smoothly goes from negative to positive.

B) In order that there can be the stability of the asymptotic state needed
there must effectively be a lower bound for the Hamiltonian density,
a bottom in the Hamiltonian density.

\section{Would our model give unacceptable effects ?: caveats}
\subsection{General risk of influence from future}
At first it seems extremely dangerous for getting very strange miraculous 
effects to introduce a probability function $P(path)$ which is sensitive to
what the path does at all times and not only in the far past.
Such a probability density would namely select the path(s) to be realized
not only with special properties (typically low entropy) in the far past
(say around the effective Big Bang time called $t_0$ above) but
also in say the near future. 
If really the path were selected -- as in our model -- also with respect to
what happens in the (near) future, such arrangement would be recognized as a 
foresight or the hand of God or a miracle. 
Nevertheless such signs of foresight are so seldom empirically 
that many people do not believe in their existence at all.
For our model to be phenomenologically viable it is therefore needed that
such effects of foresight adjustments of the path realized are extremely
weak and can in practice be ignored.

Now there is first to be taken into account that we imagine the $P$
factors coming from the inflation period, i.e. 
$\int \hat{\mathcal P}\sqrt{g}d^4x$
from this period, to be very strong so that the realized track 
is very dominantly determined to make the factor from that period large and 
from
the infinitely long time periods $t\to \pm \infty $.
The influence from the relatively short time span over which humans have
the capacity for observing the foresight or miraculous effects 
is relatively less significant.
Alone for that reason the effect is significantly suppressed.
We should remember that the special form of inflation 
-- the special inflation field and the value of it during inflation --
were first of all selected to contribute highest possible to $\log P(path)$,
so if it comes to dominate it is not so surprising again.
But at least we are able only to collect information about miracles,
foresight etc. over say of the order of thousands of years while the universe
up to now has at least existed for milliards of year, the observable effects
are in any case down by a factor a million.

Now there is, however, also an a bit more sophisticated reason for the 
suppression of such effects coming from the nature of the Standard Model.
This effect is described here below:

\subsection{Suppression of influence from future:}
First let us remark that a part in $\hat{\mathcal P}$ having the same form 
as the
kinetic term in the Lagrangian will vanish if the
free equation of motion for the field is obeyed.
If we talk about a massless particle due to some mass protection 
conservation law there will be no mass parameter, and all the non-interaction
term is already fixed.
For such a mass-protected theory a 
field-bilinear-expression can only have one form and thus must
be the same -- in reality only an $i=\sqrt{-1}$ is the difference --
form for the Lagrangian density $\mathcal L_{bil}$ and for the bilinear part 
of the
$\hat{\mathcal P}$, called $\hat{\mathcal P}_{bil}$, 
so that at least to this bilinear approximation -- no interaction 
approximation --
the $\hat{\mathcal P}_{bil}$ vanishes on shell.
According to this argument you can for mass protected particles in the free
approximation, see no effect of $P(path)$.
This result may also be seen -- may be more convincingly -- by simply using
Lorents invariance to tell that since there exists no way of having a 
Lorents invariant length of the time track of a massless particle,
we cannot possibly obtain a sensible form of the contribution to $\log P$ 
from a passing around massless particle in the approximation of the 
free path dominating.
For massive particles on the other hand there can be defined a time track 
length, since it is then possible to use the eigentime.

We may estimate the contribution coming from a Higgs particle time track 
going along.
For dimensional reasons we need a mass factor in front of the eigentime to
get a contribution to the dimensionless $\log P$.
Even if we take the mass to be the compared to fundamental scales
-- say Planck scale -- a very
%
small mass $m_H$ of the Higgs itself say, the typical life time 
$\boldsymbol{\tau}_H \approx \frac{1}{\Gamma_H}$
for a Higgs would give a contribution to $\log P$ of the order of magnitude
\begin{eqnarray}
\log P \mid _{Higgs~ flight} \approx \boldsymbol{\tau}_H m_H \approx 
\frac{m_H}{\Gamma_H}
\nonumber\\
\simeq \frac{100GeV}{\Gamma_H \sim 100MeV} \approx 10^3
~.\end{eqnarray}
If the equations of motion are not so strongly guided by what happens
at inflation time or at far future, that a factor $e^{\sim 10^3}$ in $P$ 
is not felt, this would e.g. mean that the production of a Higgs would 
somehow miraculously be prevented to the level of 
only happening once out of a priori $e^{\sim 10^3}$ expected cases.
How it would happen is not so clear, but terrorists on $LHC$ might be very
effective.

These problems deserve further investigation, but until now nobody 
saw safely the Higgs.
So a total prevention of Higgs production by mysterious foresight effects
is not out of question experimentally.

\section{Conclusion and outlook}
We have sought to build up crudely a model behind the second law of 
thermodynamics
by speculating about the properties of a ``probability weight function"
$P(path)$ which assigns to all the paths through phase space of the universe
(everything in classical thinking) a probability density relative to the phase
space density. 
We actually suggested that we could obtain a reasonable scenario without
making exceedingly specific assumptions about the form of the function of
``probability density function" $P(path)$.
Basically we actually made use of the following:
\begin{description}
\item[A)] We used a time factorized form (Locality in time),
 
\item[B)] We used that one could distinguish some macro states or classes of 
them
which were stable.

\item[C)] We used that the variation in the size of $P(path)$ were indeed huge.

\item[D)] Especially we assumed that a small or fundamental scale universe 
in inflating state would have highest $Pe^{S}$.
Very likely an inflational universe would be favoured because the scalar
fields could easily dominate $\hat{\mathcal P}$ giving 
$P \sim exp(-\int \hat{\mathcal P}\sqrt{g}d^4x)$.
\end{description}

The most concrete picture of a model unifying the second law with the other
laws would be given by a probability density with the logarithm of the form
\begin{eqnarray}
\log P(path) = 
\int \hat{\mathcal P}(\varphi, \partial_{\mu} \varphi, g_{\mu v}, 
\partial_{\rho} g_{\mu v}, \cdots ) \sqrt{g}d^4x
\end{eqnarray}
so that it is very much like an imaginary part of the complex action:
\begin{eqnarray}
\hat{\mathcal P}=2 ~Im \mathcal L_{compl.}
~.\end{eqnarray}

However, most of the structure of this expression 
$P(path)$ were not really used since the details were not so important.

Our main results were:
\begin{enumerate}
\item That in an infinite time axis model $t \in (-\infty , \infty )$ 
we could indeed obtain two ``half time axes" each of which had for itself
a second law of thermodynamics (provided that we choose the appropriate sign
on the time axis separately for the two half axes.)

\item 
 We obtain a small cosmological constant under the assumption that first the
cosmological constant could be considered part of the ``initial conditions"
and secondly that the universe develops towards a stationary asymptotic 
macro state.
This means that we need a universe developing as little as possible.

\item In an intermediate step of the argument it also meant that the density of
matter should be very low compared say to the ``fundamental density".

\item Further we got for keeping the stability that a bottom in the 
Hamiltonian density
should be called for and adjusted to exist (approximately at least).

\item We can even claim that our model had it very naturally that there 
should be
an effective Big Bang; but here it is an important point that a true Big Bang
singularity at which space-time starts is \underline{not} suggested by
our model. 
Rather the suggested scenario has the ``two sides in time" which means
that we seen from our branch have a half infinite pre-big-bang era
with \underline{falling} entropy.
Rather than any singularity we then argued for the inflationary era from the
pre-big-bang times ($t<t_0$) -- in which with our time arrow convention the
universe contracted ``inflationally" -- 
going smoothly over into the growing universe inflation for $t>t_0$.
Really we have in the time around $t_0$ -- the simulated two sided Big Bang --
just a positive curvature world developing as a de Sitter universe with the
quite smooth and natural rebounce in this case.
There is at least approximately a time reversal invariance mirror symmetry
around $t=t_0$.

This rebouncing in the inflationary stage is quite sufficient for Big Bang 
model
phenomenology because all the details on which we have phenomenological check
only goes back to the last 68 or so e-foldings in the inflational period. 
\end{enumerate}

In conclusion our picture has thus several good results:
second law of thermodynamics, small cosmological constant, bottom in 
Hamiltonian density, practical Big Bang, but in reality there is no true 
Big Bang.
There is rather a rebounce from a pre-big-bang time taking place in the 
inflation era.

With all these good features phenomenologically we have however to admit
there are also strange effects being predicted:  
With the philosophy that there is a more fundamental path dependent
probability density $P(path)$ -- even having special dependence on what happen 
today say -- there is a danger of $P$ functioning as a source of effects
which obvious to the observer, would be governed from the future
(may be even the near future).
That would look like there being a foresight.
A crude discussion leads to that a suggestive place to look for such foresight 
effects would be to investigate if the foresight would seek to prevent Higgs
particles being produced ( or perhaps instead favour Higgs production, 
depending on the sign).
Would the foresight make an accident attacking the start of $LHC$?

We think that these foresight effects that could severely threaten the
viability of our model behind the second law 
deserves further calculations as to whether the effects will be strong enough
to be really seen.

One could also imagine that because the true path realized would be guided
towards a sort of maximum in the probability density the derivatives of
$\log P(path)$ near that path will be small and that such an effect of 
derivatives at maximum being zero could lead to a reduction to zero of the
coefficients determining say how unlikely a path becomes by having a
Higgs particle produced.

Since we got the second law without much details assumed about $P$ we can
consider the present work a ``random dynamics" derivation of the second law
of thermodynamics.
That is to say we would say that second law came almost surely, almost
unavoidably.

\begin{acknowledgments}
This work was completed when one of the authors (H.B.N.) visited as a visiting 
professor to
the Yukawa Institute for Theoretical Physics (YITP), Kyoto University.
He acknowledges the member of YITP for hospitality extended to him.
The work is supported by Grant-in aids for Scientific Research 
on Priority Areas, Number of Area 763 ``Dynamics of Strings and Fields",
from the Ministry of Education of Culture, Sports, Science and Technology, 
Japan.

One of us (H.B.N.) wishes to acknowledge discussions mainly many years ago 
in a study group at the Niels Bohr Institute mainly on ``random dynamics" 
in which the $o$-operator, essentially $\log P$ appeared, especially S. E. 
Rugh,
D. Bennett and Astre Kleppe -- the latter having a homepage alluding to 
these subjects --. 
Also the discussion of the related work 
\cite{19}
in the recent Bled Conference is acknowledged.
One of the authors (M.N) acknowledges Hikaru Kawai for discussing the role of
entropy in universe in the very early stage of the present article.
\end{acknowledgments}


\begin{thebibliography}{99}
\bibitem{1}
S. Carnot, ``Ref\'{e}xions sur la Puissance Motrice du Feu et sur les Machine
Propres \`{a} Developper cette Piussance", Bechelier, Paris 1824; 
see ``The Second Law of Thermodynamics", translated and edited by 
W. F. Magie, Harper and Brothers, New York 1899, pp.3-61.

\bibitem{2}
R. Clausius, Ann. Phys. Chem., 79(1850)368-397,
500-524; ibid., 93(1854)418-506; ibid., 125(1865)353-400, 
see ``The Mechanical Theory of Heat", translated by W. R. Browne,
MacMillan and co., London, 1879;
see also W. F. Magie \cite{1}
pp.65-108

\bibitem{3}
W. Thomson, Mathematical and Physical Papers, University Press, Cambridge,
1882, Vol.1, (1848) pp.100-106; (1849)113-155;
(1851-1854, 1878)174-332;
(with J. P. Joule), (1852-1862)333-455.

\bibitem{4}
I. Prigogine: ``Thermodynamics of irreversible processes",
Interscience,
A division of Wiley and Sons,
NY London 1961; see also
J.Kirkwood, J. Chem. Phys. \underline{14}, p.180(1946)

M. Born and H. S. Green,
Proc. Roy. Soc. A 190, p.455(1947)
G. Klein and I. Prigogine,
Physica 13, p.74, 89(1957)

\bibitem{5}
S. Coleman, Nucl. Phys. \underline{B 307} 867(1988)

\bibitem{6}
S. Coleman, Nucl. Phys. \underline{B 310} 643(1988)

\bibitem{7}
T. Banks, Nucl. Phys. \underline{B 309} 493(1988)

\bibitem{8}
S. W. Hawking, Phy. Lett. \underline{134B} 403(1984)

\bibitem{9}
E. I. Guendelman and A. B. Kaganovich, 
``Dynamical measure and field theory models free of the cosmological constant
problem", gr-qc/9905029(1999), Phys. Rev. \underline{D53}, 7020(1996);
see also in: Proceedings of the third Alexander Friedmann International 
Seminar on Gravitation and Cosmology, edited by Yu. N. Gnedin, A. A. Grib, 
V. N. Nostepanenko (Friedmann Laboratory Publishing, St. Petersburg, 1995);
Phys. Rev. \underline{D55}, 5970(1997);
Mod. Phys. Lett. \underline{A12}, 2421(1997);
Phys. Rev. \underline{D56}, 3548(1997);
Hadronic Journal \underline{21}, 19(1998);
Mod. Phys. Lett. \underline{A13}, 1583(1998).

\bibitem{Weinberg} S. Weinberg, ``The Cosmological Constant Problem'', 
Rev. Mod. Phys. {\bf 61},1-123(1989).

\bibitem{10}
N. C. Tsamis and R. Woodard, Phys. Lett. \underline{B301}, 351-357(1993)

\bibitem{11}
S. W. Hawking, Phys. Rev. \underline{D47}, 5342-5356(1993),
gr-qc/9301017

\bibitem{12}
S. W. Hawking, Phys. Rev. \underline{D32}, 2489(1985)

\bibitem{13}
J. B. Hartle and S. W. Hawking, Phys. Rev. \underline{D28}, 2960-2975(1983)

\bibitem{14}
L. Boltzmann, Ann. Physik
\underline{60}, 392(1897), translated in S. G. Brush, 
Kinetic Theory (Pergamon Press, New York, 1965)
Quoted in: James B. Hartle, Quantum Pasts and the Utility of History, 
gr-qc/9712001, Talk present at The Nobel Symposium: Modern Studies of Basic 
Quantum Concepts and Phenomena, Gimo, Sweden, June 13-17, 1997.

\bibitem{15}
S. W. Hawking and Thomas Hertog, Why Does Inflation Start at the Top of the Hill?,
hep-th/0204212; S. W. Hawking and Don Page, Nucl. Phys. \underline{B298}:789(1988)

\bibitem{16}
H. B. Nielsen and S. E. Rugh,
``Arrows of time and Hawking's no-boundary proposal",
Neils Bohr Institute Activity Report 1995,
Murray Gell-Mann and James B. Hartle, Time Symmetry and Asymmetry in
Quantum Mechanics and Quantum Cosmology, arXiv: gr-qc/9304023.

\bibitem{17}
P. J. Steinhardt, Neil Turok,
hep-th/0111098, Phys. Rev. \underline{D65}: 126003, 2002

\bibitem{18}
Masafumi Fukuma, Hikaru Kawai, Masao Ninomiya, hep-th/0307061,
Int. J. Mod. Phys. \underline{A19}: 4367-4386, 2004.

\bibitem{19}
H. B. Nielsen and M. Ninomiya, 
Compactified Time and Likely Entropy -- World Inside a Time Machine: 
Closed time-like Curve.-- 
To appear in Proceedings for ``What comes beyond the Standard Model?"
held in Bled, Slovenia, 18th-29th of July 2005. YITP-05-38 and OIQP-05-06.

\bibitem{inflation} A.H. Guth, The Inflationary Universe: A possible Solution
to the Horizon and Flatness Problems, Phys. Rev. {\bf D23},347 (1981), A.D. 
Linde, A New Inflationary Universe Scenario: A possible Solution of the 
Horizon, Flatness, Homogeneity, Isotropy, and Primordial Monopole Problems, 
Phys. Lett. {\bf B108} 389(1982); A. Albrecht and P. Steinhardt,Phys. Rev. 
Lett.{\bf 48},1220 (1982); A. A. Starobinski, Phys.Lett. {\bf B91},99(1980);
A. D. Linde, Phys. Lett. {\bf B129},177 (1983); D. La and D. J. Steinhardt,
Phys. Rev. Lett. {\bf 62},376 (1989). 
\end{thebibliography}
\end{document}